\documentclass[12pt,preprint]{aastex}

\newcommand{\be}{\begin{equation}}
\newcommand{\ee}{\end{equation}}
\newcommand{\etal}{et\ al.\ }

\begin{document}

\title{Cosmic Microwave Background Anisotropy Measurement From Python V}
\author{K. Coble\altaffilmark{1,2}, S. Dodelson\altaffilmark{3,1},
M. Dragovan\altaffilmark{4}, K. Ganga\altaffilmark{5},
L. Knox\altaffilmark{6}, J. Kovac\altaffilmark{1}, B. Ratra\altaffilmark{7}, and T. Souradeep\altaffilmark{7,8}}

\altaffiltext{1}{Enrico Fermi Institute, University of Chicago, 5640 South Ellis Ave., Chicago, IL 60637, USA, coble@hyde.uchicago.edu.}
\altaffiltext{2}{Adler Planetarium and Astronomy Museum, 1300 South Lake Shore Drive, Chicago, IL 60605, USA.}
\altaffiltext{3}{Fermi National Accelerator Laboratory, Box 500 MS-209, Batavia, IL 60510, USA.}
\altaffiltext{4}{Jet Propulsion Laboratory, California Institute of Technology, 4800 Oak Grove Drive, 169-506, Pasadena, CA 91109, USA.}
\altaffiltext{5}{Infrared Processing and Analysis Center, California Institute of Technology, Mail Code 100-22, Pasadena, CA 91125, USA.}
\altaffiltext{6}{Department of Physics, University of California, One Shields Avenue, Davis, CA 95616, USA.}
\altaffiltext{7}{Department of Physics, Kansas State University, 116 Cardwell Hall, Manhattan, KS 66506, USA.}
\altaffiltext{8}{IUCAA, Post Bag 4, Ganeshkhind, Pune 411007, India.}


\begin{abstract}

We analyze observations of the microwave sky made with the Python
experiment in its fifth year of operation at the Amundsen-Scott South
Pole Station in Antarctica. After modeling the noise and constructing a map,
we extract the cosmic signal from the data. We simultaneously
estimate the angular power spectrum in eight bands ranging from large
($\ell \sim 40$) to small ($\ell \sim 260$) angular scales, with power detected
in the first six bands. There is a significant
rise in the power spectrum from large to smaller ($\ell \sim 200$) scales, 
consistent with that expected from acoustic oscillations in the early
Universe. We compare this Python V map to a map made from data taken
in the third year of Python. Python III observations were made at a
frequency of 90 GHz and covered a subset of the region of the sky
covered by Python V observations, which were made at 40 GHz.  Good
agreement is obtained both visually (with a filtered version of the
map) and via a likelihood ratio test.

\end{abstract}



\section{INTRODUCTION}

Since the detection of anisotropy in the cosmic microwave background
by the COBE satellite, many experiments have measured the angular
power spectrum at degree and sub-degree angular scales (e.g.,
Netterfield et al. 2002; Halverson et al. 2002; Lee et al. 2001;
Miller et al. 1999).  The Python V data set has sufficient sky coverage to
probe the smallest scales to which COBE was sensitive, while having a
small enough beam to detect the rise in the angular power spectrum to
degree angular scales, providing a link in $\ell$-space between COBE and
other recent measurements.

Python V is the latest of the Python experiments at the South
Pole. Dragovan et al. (1994), Ruhl et al. (1995), and Platt et al.
(1997) describe Python I--III and Rocha, et al. (1999) derive constraints
on cosmological parameters from these data. Kovac et al. (1997)
describe the Python IV results.

The Python V experiment, observations, and data reduction are
described in Coble et al. (1999). In that paper, we
analyzed individual modulations of the data.  The modulations
can be thought of as filters which have little sensitivity to some of
the contaminants in the time stream.  For example, they have no
sensitivity to gradients, which should get a large contribution from
the atmosphere and from the ground shield.
The modulation approach also provided a rapid
means of compressing a large amount of data (19 Gbytes) into a more
manageable size. Measurements of anisotropy were reported for eight
different modulations of the sky signal; the results indicated a sharp
rise in the power spectrum.

In this paper we find the constraints on the power spectrum due to all
of the modulations simultaneously.  We use the modulations as our
starting point, rather than the time stream, to take advantage of the
contaminant filtering and data compression.  We extend the analysis of
Coble et al. (1999) by accounting for the correlations (in both signal
and noise) between different modulations.  From the modulations we
find the best-fit map and its associated noise covariance.  From this
map and its associated covariance matrix we estimate the power
spectrum simultaneously in eight bands.

In $\S$ 2 we briefly review the instrument and the data set.
In $\S$ 3 we discuss
the estimation of the noise matrix. In $\S$ 4 we describe how to
use this matrix to construct a map and a noise matrix for the map.
This map is used in $\S$ 5 to estimate the angular power spectrum
in eight bands.  In $\S$ 6 we check the power spectrum derived from
the map with the power spectrum derived directly from the modulated
data.  In $\S$ 7 we compare the 40 GHz Python V data with the 90
GHz Python III data (Platt et al. 1997), which covered a subset of the
region of the sky covered by Python V.  We find good agreement between
the two observations in the region of overlap, providing a
valuable consistency check. This is another
indication of a lack of significant foreground contamination (see also our
estimates in Ganga et al. 2002). We conclude in $\S$ 8.


\section{INSTRUMENT AND DATA}

We begin with a brief review of the Python V instrument and data,
emphasizing the terminology used to describe the different subsets of
data. More detailed descriptions of the instrument can be found
in Coble (1999) and Alvarez (1996). A more detailed description
of the Python V data set can be found in Coble (1999).

The receiver consists of two focal-plane feeds, each with a
single 37-45 GHz HEMT amplifier. The two focal-plane feeds of the
receiver correspond to two beams at the same declination separated by
2.80$^\circ$ on the sky. Each of the two feeds is split into two
frequency channels near 40 GHz, yielding a total of four data channels.
The receiver is mounted on a 0.75 m-diameter off-axis parabolic
telescope, which is surrounded by a large 12-panel ground shield.
The instrument was calibrated using thermal loads for the DC calibration;
the overall uncertainty in the calibration of the data set is
estimated to be (+15\%, -12\%) in $\Delta T$. The combined absolute and
relative pointing uncertainty is estimated to be $0.15^{\circ}$, 
as determined by measurements of the Moon and the Carina nebula
($\alpha=10.73^{\rm h}, \delta=-59.65 ^{\circ}$).
The Python V beam is well approximated by an asymmetric Gaussian of FWHM
$0.91^{+0.03}_{-0.01} {\rm deg} \times 1.02^{+0.03}_{-0.01}$ deg ($az \times el$),
as determined from scans of the Carina nebula and the Moon.
Given this beam size uncertainty of approximately $0.015^\circ$,
the band power can move roughly by a factor of
$\exp ( \pm l(0.425)(0.015)(\pi/180) )$, only a $3\%$
effect at $l=200$.

Python V observations were taken from November 1996
through February 1997. Two regions of sky were observed:  the Python V main
field, a $7.5^{\circ} \times 67.7^{\circ}$ region of sky centered at
$\alpha=23.18^{\rm h}$, $\delta=-48.58^{\circ}$ (J2000) which includes 
fields measured during the previous four seasons of Python observations and
a $3.0^{\circ} \times 30.0^{\circ}$ region of sky centered
at $\alpha=3.00^{\rm h}$, $\delta=-62.01^{\circ}$ (J2000), which
encompasses the region observed with the ACME
telescope (Gundersen et al. 1995).
The total sky coverage for the Python V regions is 598 deg$^2$.

\begin{figure}
\epsscale{1.0}
\plotone{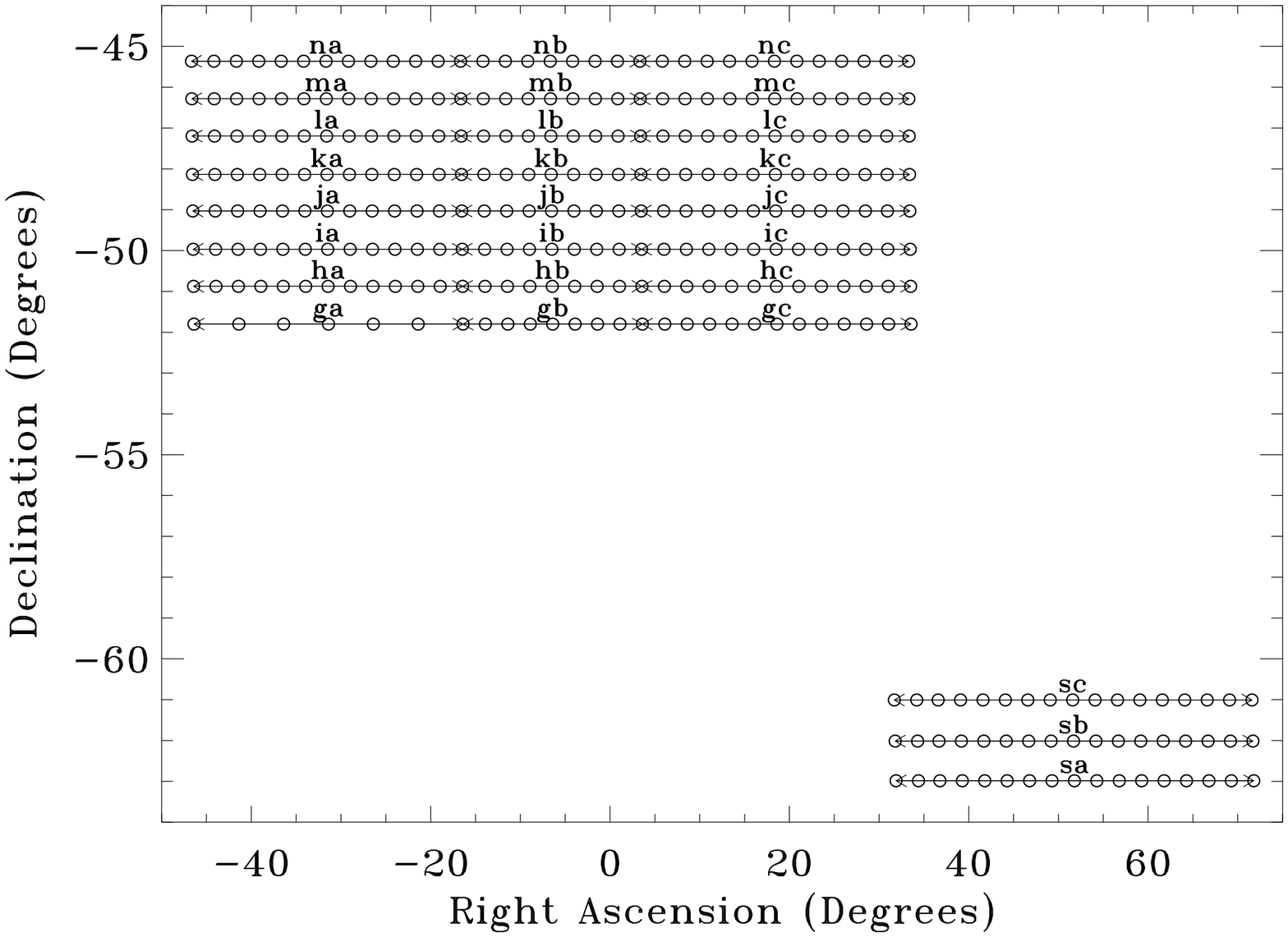}
\caption{Observing sets. Each set is observed for
approximately a day (leaving $\sim$ 13 hours of good data) before moving
on to the next set. Neighboring sets in the main Python V region overlap
by 1 field. Circles represent the fields and arrows point to the
end fields in each set.
In addition to the sets shown, sets ib, jb, kb, and lb were
observed with a scan pattern of 5 fields per file.
\label{fig1}}
\end{figure}

Both Python V regions are observed with a grid spacing of 0.92$^{\circ}$
in elevation and 2.5$^{\circ}$ in right ascension, in 345 effective
{\it field}s. There are 309 unique field positions, but some positions
are observed at different times of the observing season and are thus
counted as different fields for analysis purposes.
The telescope is positioned on one of the fields and the chopper
smoothly scans the beams 17$^{\circ}$ in azimuth in a nearly triangular wave
pattern at 5.1 Hz. One {\it cycle} corresponds to all of the data
taken in one back-and-forth scan along the sky. A cycle consists of
$128$ {\it samples} along the sky in the given field. A {\it stare} is
$164$ cycles, again centered on the same spot on the sky. One data
{\it file} consists of roughly ten stares, at adjacent fields on the
sky. Typically, a file corresponds to data taken over five to ten
minutes (depending on how many stares it contains).  The telescope
remains on this {\it set} of fields for roughly $13$ hours, so any set
of fields is typically observed in about one hundred consecutive files.
This terminology is summarized in Table 1 and illustrated
in Figure \ref{fig1}.

\placetable{tbl1}

\placefigure{fig1}

In software, the data are modulated such that the
{\it spatial} responses are cosines apodized with a Hann
window. In order to take advantage of the large sky coverage of Python V,
which allows us to probe large angular scales, we also use an additional
cosine modulation which was not apodized by a Hann window.
Data taken during the right
and left-going portions of the chopper cycle are modulated separately, to
allow for cross-checks of the data. Sine modulations are not used in the
analysis because they are anti-symmetric and are thus sensitive to
gradients.
The modulated data in a given stare is a linear
combination of the samples:
\begin{equation}
D_{ima} = \sum_{s=1}^{128} M_{ms} d_{isa}.
\label{eq:mods1}
\end{equation}
The index $i=1,\ldots,690$ labels the field and feed; $m=1,\ldots,8$ labels the
modulation; $a=1,\ldots,N_f$ labels the file which looks at a given
set of fields; $s=1,\ldots,128$ indexes the sample number; $d$ is the
unmodulated data which has been co-added over all cycles in a stare.

A chopper synchronous offset, due to differing amounts of spillover,
is removed from each data file by subtracting the average of all
stares in a file. This is not just a DC offset; there is an offset
removed for each modulation. The chopper synchronous offset is
discussed in detail in Coble (1999), but typical values are 100 to 200
$\mu {\rm K}$ for each modulation, stable over a timescale of $\sim 5$ files.

When the data are binned in terrestrial
azimuth, a periodic signal due to the 12 panels of the ground shield
is evident, especially in the lower-$\ell$ modulations. This signal of
period 30$^{\circ}$ is fit for an amplitude and is subtracted.
Removal of the ground shield offset
has less than 4\% effect on the final angular power spectrum
because when the data are binned in RA, the effect
averages out. The ground shield offset is discussed in detail in
Coble (1999), but typical values for the signal amplitude are
less than 100 $\mu {\rm K}$.

Both the chopper synchronous offset and the ground shield offset
subtractions are accounted for by adding a constraint matrix,
{$\bf C^C$}, to the noise matrix (Bond, Jaffe, \& Knox 1998). The precise form
of these matrices is given in Coble (1999). Their impact on the final
result is minimal since they serve to remove only a handful of modes
from the analysis. In the Coble \etal (1999) analysis, the chopper
synchronous offset
and the ground shield signal were removed from the data, but the ground shield
constraint matrix was not included in the analysis. The constraint
matrix for the chopper synchronous offset was included in that analysis.

After the data have been modulated and offsets removed, the right- and
left-going data, which have been properly phased, are co-added, as are
data from channels which observe the same points on the sky.
Data pointing at a field $i$ from all files are averaged to form
\begin{equation}
D_{im} = {1\over N_f} \sum_{a=1}^{N_f} D_{ima}.
\end{equation}
This final data vector has $5520$ ($=345$ fields $\times$ 2 feeds
$\times$ 8 modulations) components.  The next section describes our
modeling of the noise properties of these data.


\section{NOISE MODEL}

Accurate modeling of the noise is often one of the most difficult
tasks in CMB analysis. The noise model we develop below enables
us to estimate the angular power spectrum in
eight bands {\it simultaneously}. See Coble (1999) for a
more detailed discussion of the noise modeling of the Python V data.
In Coble et al. (1999), we modeled the noise only for
individual modulations. The noise model described here also models
the cross-modulation terms, allowing us to include cross-modulation
correlations in the power spectrum analysis. The noise level for the
Python V data is $\lesssim 1 {\rm mK s}^{1/2}$.

Our noise model assumes that the covariance between fields taken with
different sets of files is negligible (as in Coble et al. 1999)
because of the chopper offset removal and because of the long time
between measurements. An analysis comparing the noise estimated
on different timescales indicates that Python V noise is dominated by
detector noise and is Gaussian.

Since many different files look at the same field on the sky, there is
a simple way to estimate the noise covariance matrix.  We first
estimate the noise matrix via:
\begin{equation}
\hat C^N_{ijmm^{\prime}}
= {1\over N_f} \sum_{a=1}^{N_f} 
( D_{ima} - D_{im} ) 
( D_{jm'a} - D_{jm'} )
\label{eq:naive}
\end{equation}
where again $i,j$ index the different fields, $m,m'$ the eight
modulations, and we sum over all $N_f$ files which observe the fields
of interest. However, since there are typically only 100 files
for each field, the sample variance on the noise estimate
is $\sim 1/(100)^{1/2}$, or 10\%, which will
severely bias estimates of band power. Hence we do not
use this naive estimator.

To obtain a better estimate of the noise, in Coble et al. (1999),
we averaged the
variances for each set of files and then scaled the off-diagonal
elements of the covariance to the average variance in a given set based
on a model derived from the entire Python V data set. In that paper,
$C^N_{ijmm'}$ was computed for each individual modulation,
i.e., with $m=m'$ only.
Several consistency checks were performed showing that
the final noise model for
the single modulation analysis was a good one.

In this cross-modulation analysis, we initially extended the method
in Coble et al. (1999) to account for cross-modulation terms in
$C^N_{ijmm'}$, i.e., terms with $m \neq m'$.
To test this noise model, we constructed $\chi^2={\bf D}^t({\bf
C^N}+{\bf C^C})^{-1}{\bf D}$, for each observing set (which typically
includes of order ten fields observed $\sim 100$ times each).  There
is very little CMB signal in any one set, so we expect $\chi^2/{\rm dof}$ to
be close to one.  The results fail this $\chi^2$ test,
indicating that a better model of the cross-modulation noise is necessary.

To go beyond the initial estimators, we assume
the cross-modulation noise matrix factors as
\begin{equation}
C^N_{ijmm^{\prime}}=C^M_{mm^{\prime}} C^F_{ij},
\label{eq:cnijmm}
\end{equation}
where {$\bf C^M$} describes the cross-modulation correlations and {$\bf C^F$}
the field-field correlations. The cross-modulation correlations
are derived from the sample-space covariance matrix {$\bf C^S$}: 
\begin{equation}
C^M_{mm^{\prime}}= \sum_{s,s'=1}^{128} M_{ms} C^S_{ss'} M_{m's'}.
\label{eq:cnmm}
\end{equation}
The matrix ${\bf C^S}$ describes the noise in the timestream as a
function of chopper sample $s$.  To clarify, ${\bf C^S}$ is a 128
$\times$ 128 matrix, ${\bf C^M}$ is a $8\times 8$ matrix, and ${\bf
C^F}$ is a 690 $\times$ 690 matrix for Python V.  Models for ${\bf
C^S}$ and ${\bf C^F}$ are needed in order to construct ${\bf C^N}$.

\begin{figure}
\epsscale{1.0}
\plotone{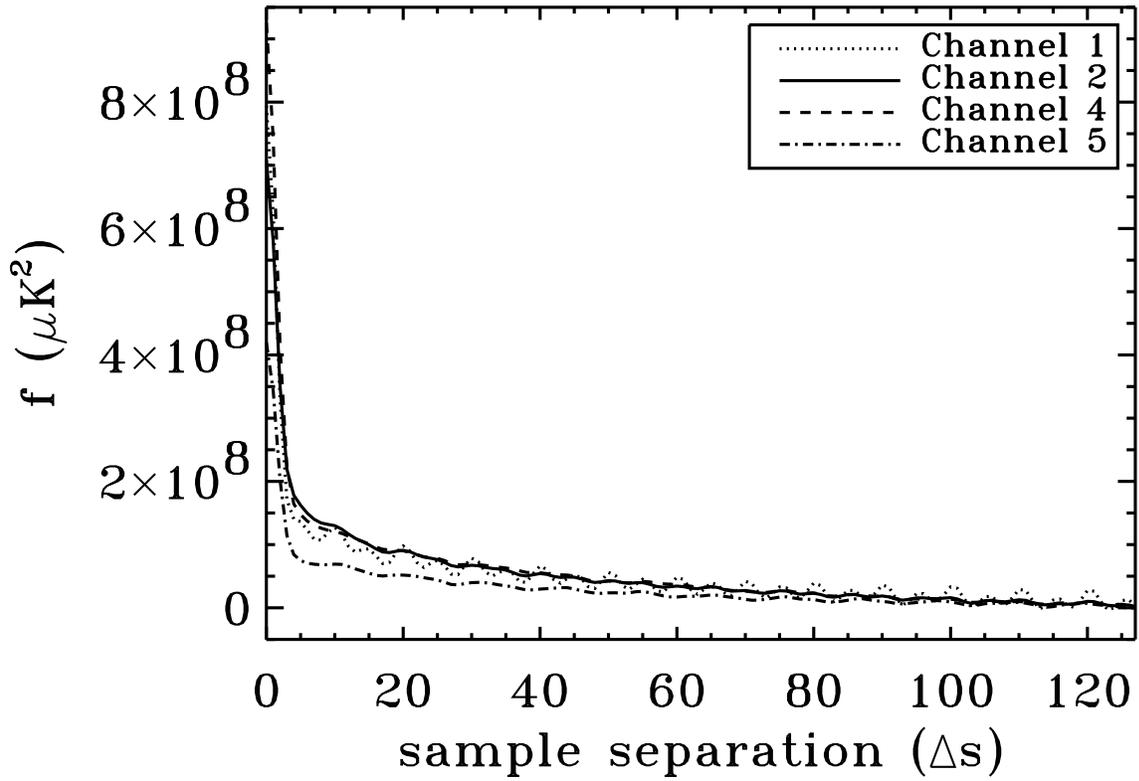}
\caption{Correlation as a function of sample separation,
$f(\Delta s)$, for all four data channels in one of the sets of files.
The noise is a combination of atmospheric and
instrumental noise. Channel 5 is our most
sensitive channel. Channel 3 is a dark channel.
These noise levels correspond to $\lesssim 1 {\rm mK s}^{1/2}$.
\label{fig2}}
\end{figure}

If we assume that ${\bf C^S}$ depends only on
chopper sample separation $\Delta s=s-s^{\prime}$,
it can be computed from the following function:
\begin{equation}
f(\Delta s)={1\over N_S} \sum_{s}d_{s}d_{s+\Delta s}
\label{eq:fdeltas}
\end{equation}
where $N_S$ is the number of samples and $d_s$ is the unmodulated data.
For example, the $C^S_{12}$ component is given
by $f(\Delta s = 1)$.
In order to compute $f(\Delta s)$, a chopper synchronous offset is
first subtracted from the raw data. Then
$f(\Delta s)$ is calculated for each
channel, cycle and stare in a file. $f(\Delta s)$ is
then averaged over cycles, stares, and files.
Figure \ref{fig2} shows $f(\Delta s)$ for each channel
in one of the sets.

\placefigure{fig2}

\begin{figure}
\epsscale{0.7}
\plotone{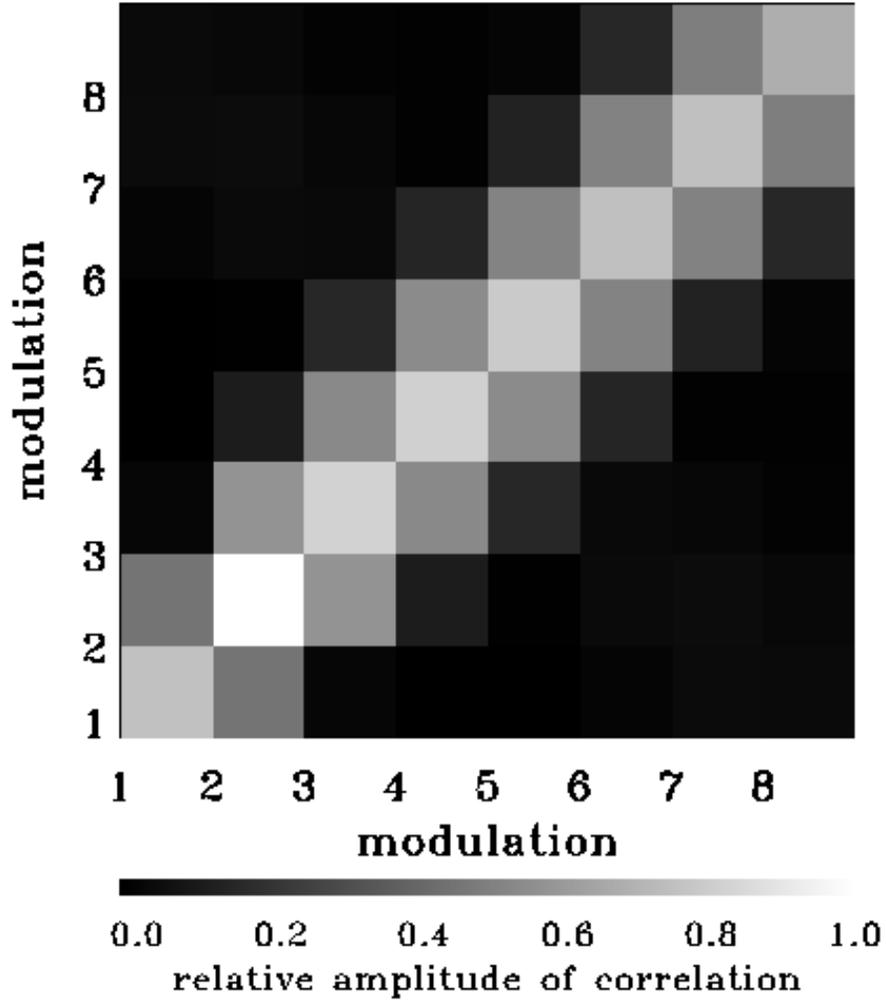}
\caption{${\bf C^M}$ for one
set and feed. Elements that are more than two modulations
apart are relatively uncorrelated.
\label{fig3}}
\end{figure}

With this model for the sample correlation function, it
is now straightforward to compute ${\bf C^M}$ for each set and channel
following equation (\ref{eq:cnmm}). ${\bf C^M}$ matrices for
channels which look at the same point on the sky and for
right and left-going chopper data are averaged, yielding
${\bf C^M}$ matrices for both feeds in each set.
As an example, ${\bf C^M}$ for one set and feed is shown
in Figure \ref{fig3}.

\placefigure{fig3}

In order to get a simple form for ${\bf C^F}$, the field
correlation matrix, we ignore the correlations between the
two feeds and assume the correlation between
fields $i$ and $j$ come only from the chopper offset subtraction.  We
investigated several similar noise models and found that these
assumptions do not change the single modulation angular power spectrum
significantly, so we assume ${\bf C^F}$ is of this form for the
cross-modulation analysis.

\begin{figure}
\epsscale{1.0}
\plotone{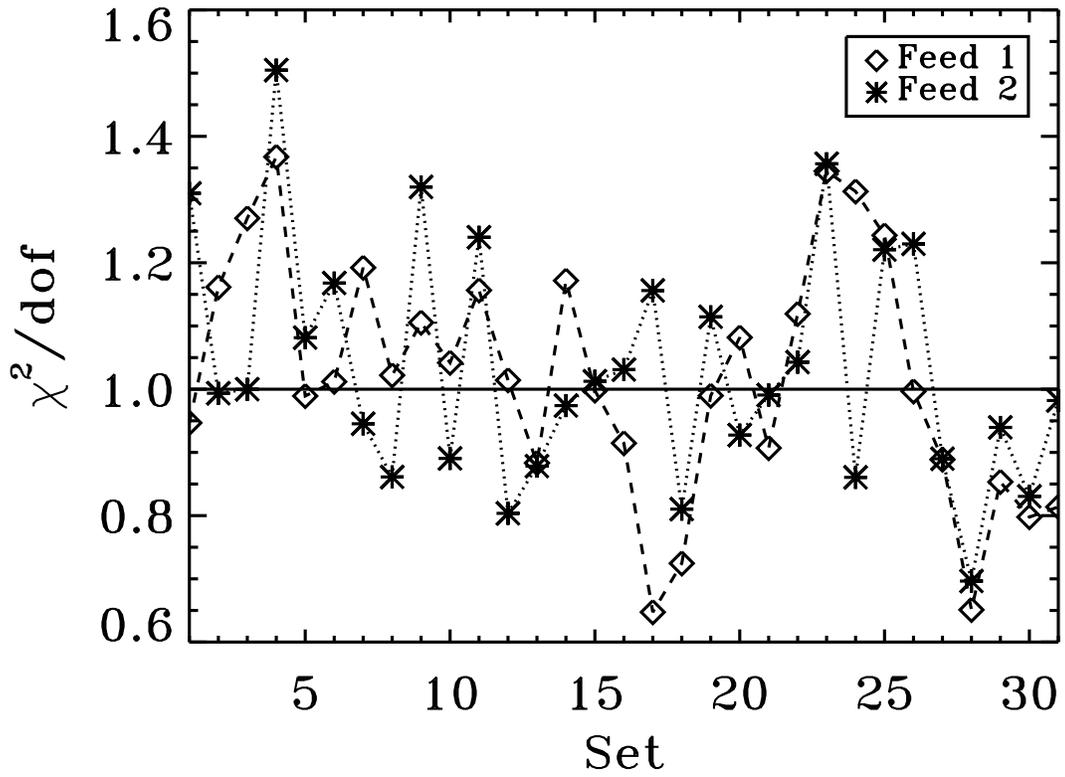}
\caption{$\chi^2$/dof for each set using the final
noise matrix.
Typical number of dof is 96 (= 13 stares $\times$ 8 modulations less 8 constraints),
though some are much smaller.
The $\chi^2$/dof close to 1 indicates that the final cross-modulation
noise model is a good one.
\label{fig4}}
\end{figure}

Finally, since this noise model is derived from sample to sample 
fluctuations, it is larger
than the corresponding noise derived from the co-added data by a factor
of $\sim 10^4$, so $C^N_{ijmm^{\prime}}$ must be normalized to the
variance in the co-added data for each set. Since ${\bf C^M}$
accounts for the relative normalization of all of the modulations,
$C^N_{ijmm^{\prime}}$ must be normalized to the
variance in only one modulation of the co-added data for each set.
We normalize to modulation 8 because we expect the higher order 
modulations to be
least affected by the ground shield.
Figure \ref{fig4} shows the $\chi^2$/dof for each set
using the final cross-modulation noise model, indicating
a good final noise model.

\placefigure{fig4}

As another check on the noise matrix used in the cross-modulation
analysis, single modulation band powers were computed using
the $C^N_{ijmm}$ components of $C^N_{ijmm^{\prime}}$. These are
consistent with the band powers given in Coble et al. (1999).


\section{MAPS}

We want to estimate the power spectrum from the Python V
data. Ideally, given the non-circular beam, this should be done
directly from the modulated data.  This would require us to form the
likelihood function from the covariance matrix for the $N=5520$ data
points. Inversion or Cholesky decomposition of matrices are $N^3$
processes so computational demands are significantly alleviated by
creating a map (with $N=1666$ at highest resolution) from the
modulated data and then estimating the power spectrum from the map.
This technique was used in an analysis of the MSAM-I experiment
(Wilson et al. 2000), for which the power spectrum estimated using the
map is consistent with the power spectrum estimated directly from the
modulated data. The inversion problem of map making and the circular
beam assumption used in it does call for cross-checks and verification
against likelihood analyses of the modulated data. The results of such
tests are summarized in $\S$ 6.

The data can be expressed as:
\begin{equation}
{\bf D} =  {\bf M}{\bf T} + {\bf n}
\label{eq:tdef}
\end{equation}
with noise covariance matrix $\langle nn \rangle = {\bf N} = {\bf C^N}+{\bf C^C}$.
As mentioned above, the data vector $\bf D$ has $5520$ elements. 
The matrix ${\bf M}$ describes the experimental processing of the
underlying temperature field; it is equal to the modulations
with an index corresponding
to each pixel at which we estimate the temperature $\bf T$.
Given the modeling of the data as in equation (\ref{eq:tdef}),
the  minimum variance estimator for $\bf T$ is
\begin{equation}
\hat T = \tilde {\bf N} {\bf M} {\bf N^{-1}} {\bf D}.
\label{eq:lincom}
\end{equation}
This estimator will be distributed around the
true temperature due to noise,
where ${\bf \tilde N}$, the noise covariance
matrix for the map, is given by
\begin{equation}
{\bf \tilde N} 
\equiv \langle (\hat T - {\bf T})(\hat T - {\bf T}) \rangle = 
{\bf \left(  M^T  N^{-1} M \right)^{-1}}.
\label{eq:cn_map}
\end{equation}

The inversion in equation~(\ref{eq:cn_map}) is singular, so it is performed
via Singular Value Decomposition. It is obvious which modes are
singular and should be neglected.  We have tested various thresholds
and found no change in the results.  The pixels in the map are $30'$
in RA which corresponds to about $20'$ on the sky. Coarser grids gave
similar results for the band powers; as we will see, there is little
sensitivity to modes with $\ell>200$, so $20'$ (a third of the beam size)
is more than adequate.

Another advantage of the map basis is that the theory covariance
matrix is simple to compute.  In the map basis, the theory covariance
matrix simplifies to
\begin{equation}
{C^T_{ij}} = <T_i T_j > = \sum_\ell {2\ell+1\over 4\pi} 
P_\ell(\cos\theta_{ij}) e^{-\ell^2\sigma^2} C_\ell,
\label{eq:ct_map}
\end{equation}
where $i$ and $j$ now refer to map pixels, $P_\ell$ is a Legendre polynomial, 
and $\theta_{ij}$ is the angular separation between points. We take $\sigma =
(\sigma_{az} \sigma_{el})^{1/2} = 0.425 \times 0.96^{\circ}$.
Taking the beam to be circular will not change the band
powers significantly (see $\S$ 6). From equation~(\ref{eq:ct_map}), the window
functions in the map basis only depend on the angular separation
$\theta_{ij}$ and not on any of the details of the observing strategy.
This is a smaller basis than that used in the analysis of the modulated
data which accounts for beam non-circularity, described in $\S$ 6.
Indeed, one way to think of a map is that it is the linear combination
of the data for which the signal (and therefore its covariance) is
nearly independent of the specific experimental observing strategy.  The
noise covariance for the map (eq. [\ref{eq:cn_map}]) accounts for
all of the experimental processing and the constraints.

\begin{figure}
\epsscale{0.5}
\plotone{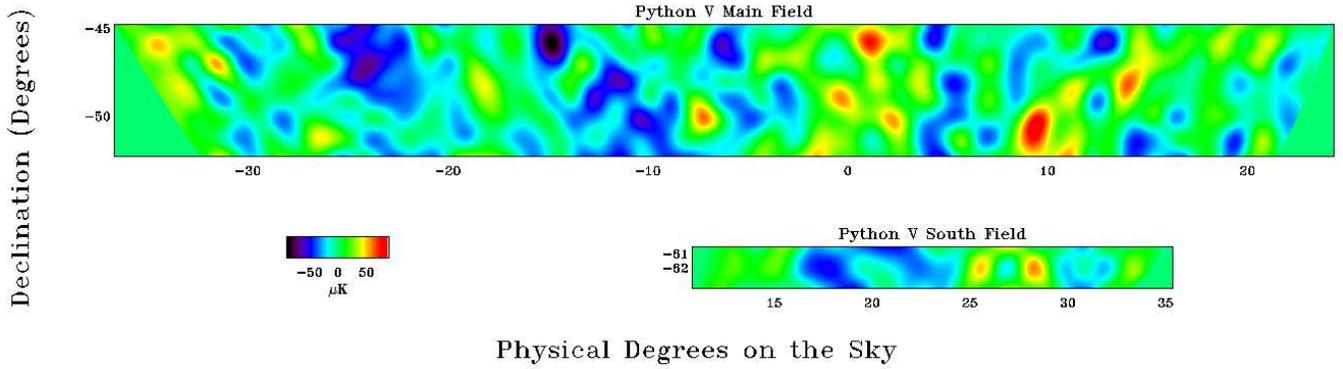}
\caption{Wiener-filtered CMB map for the Main and South Python V regions. The maps are plotted with the same size and temperature scales. The unfiltered map was used for power spectrum estimation.
\label{fig5}}
\end{figure}

Although we are primarily interested in the map as a vehicle on the
road to the power spectrum, it can also be Wiener-filtered to produce
a realistic image of the sky. Wiener-filtered maps of both of the Python
V regions are shown in Figure \ref{fig5}.  We use the unfiltered
map for power spectrum
estimation. The map serves another useful function apart from its use
for the power spectrum. One can use maps to compare different data sets 
that were
processed in completely different manners.  In $\S$ 7 we present a
visual comparison of Python III and Python V. First, though, let us
compute the power spectrum.

\placefigure{fig5}


\section{ANGULAR POWER SPECTRUM FROM THE MAP}

\def\cd{ {\cal C}}
\def\cl{ {\cal C}_\ell}
\def\C{ {\cal C}}
We now use the map to estimate the CMB anisotropy power spectrum.  Because
the observations are far short of full sky coverage, we
cannot determine individual $C_\ell$'s.  Instead, 
we parameterize the theory covariance
matrix, {$\bf C^T$}, with the power
spectrum, $\cl \equiv \ell(\ell+1) C_\ell/(2\pi)$,  
broken into bands of $\C_\ell$, denoted by $a$
\begin{equation}
\cl = \sum_a \chi_{_{a(\ell)}} \C_a;
\chi_{_{a(\ell)}}  = \left\{ 
\begin{array}{r@{\quad:\quad}l}
1 & \ell_{min}(a) < \ell < \ell_{max}(a) \\
0 & {\rm otherwise}
\end{array} 
\right.
\label{eq:cl}
\end{equation}
so that within each band $a$, $\cl=\C_a={\rm constant}$
and $\ell_{min}(a)$ and $\ell_{max}(a)$ delimit the range of band
$a$. We use eight contiguous bands of equal width, as given in
Table 2.

Since the CMB anisotropy appears 
to be Gaussian on the angular scales probed by the Python V experiment
(Park et al. 2001; Wu et al. 2001; Shandarin et al. 2002),
we can in principle use the theory covariance matrix for the
map (eq.~[\ref{eq:ct_map}]) together with the map noise
matrix, $\tilde N$, and the pixelized map data, $\bf T$,
to form the full likelihood function:
\begin{equation}
{\cal{L}}={(2\pi)^{-N/2}}
{{\rm det}({\bf C})}^{-1/2}{\rm exp}{(-\chi^{2}/2)}
\label{eq:like}
\end{equation}
where $\chi^{2} = \bf T^t {\bf C^{-1}} \bf T $ and ${\bf C}={\bf
C^{T}+ \tilde N}$.  We can then find the $\C_a$ which maximize it by
conducting a direct, grid-based search in the full eight-dimensional
parameter space.  In practice, this is of course unfeasible because it
would require of order ten likelihood evaluations in every dimension
of parameter space. The likelihood function computation requires an
inversion and a determinant of a large matrix (in our finest
pixelization, $1666\times 1666$), so it is certainly impractical to
attempt this $10^8$ times.

Instead, we use the quadratic estimator (Bond, Jaffe, \& Knox 1998; Tegmark 1997) 
to find the maximum likelihood band powers and their errors. Defining 
\be
A_{a,ij} \equiv \left[{\bf C^{-1}} {\partial C_T \over \partial \C_a} 
{\bf C^{-1}}\right]_{ij}
\ee
where {\bf $C$} is the full theory plus noise covariance matrix,
the Fisher Matrix which describes the errors is 
\be
F_{ab} = 
{1\over 2}{\rm Tr}\left[{\partial C_T \over \partial \C_a}   A_b
                        \right]
\ee
and the quadratic estimator is
\be
\hat \C_a = \C_a^{(0)} + {1\over 2} F^{-1}_{ab}
\Big( \bf T A_b \bf T - {\rm Tr}\left[ {\bf C^{-1}} 
{\partial C_T \over \partial \C_b}   \right]\Big)
.\ee
We start from a flat spectrum (e.g. all $\C_a^{(0)} = 1000\ \mu {\rm K}^2$)
and iterate four times. Convergence to well within the
size of the error bars is usually reached by the second
iteration.

The band temperature $\Delta T_\ell (= \cl^{1/2})$
results of the likelihood analysis are shown in Figure
\ref{fig6} and given in Table 2. The $\Delta T_\ell$
error bars here account only for the statistical uncertainties, and in
particular, do not account for the calibration or beam uncertainties.
The Fisher matrix is given in Table 3. We emphasize that the
values of the angular power spectrum differ from those in Coble
et al. (1999) because we are including more information in this
analysis: the cross-modulation correlations. Again, single modulation
band powers computed using the $C^N_{ijmm}$ components
of $C^N_{ijmm^{\prime}}$ are
consistent with the band powers given in Coble et al. (1999).

\placefigure{fig6}

\placetable{tbl2}

\placetable{tbl3}


\section{Comparison of Band Temperatures from Modulated Data and Map}

The map has a much smaller basis ($N=1666$ at the finest pixelization) than 
the modulated data ($N=5520$) and hence allows for speedier likelihood 
analysis. However, map making is an extra step that needs verification. In 
particular, the map analysis has to assume a circular beam when constructing 
the theory covariance matrix {$\bf C^T$}. 

In the modulated data basis, the beam corresponds unambiguously to the
measured beam response function. The non-circular (elliptical) Python V
beam is an additional complication for {$\bf C^T$} computations.
Souradeep \& Ratra (2001) develop computationally rapid methods for
computing {$\bf C^T$} for experiments with non-circular beams. The constant 
elevation scans of the Python V experiment allow us to exactly incorporate 
the effects of beam non-circularity, without recourse to any approximation.

The larger size of the modulated data basis makes the 8 band likelihood 
analysis described in $\S$ 5 computationally expensive. We therefore choose 
to compare the map basis and modulated data basis results using a simplified 
analysis that accounts for the cross-correlation between modulations in a 
limited manner. This likelihood analysis estimates the band temperatures in
each of the 8 $\ell$-space bins while holding fixed the other 7 band
temperatures at the central values obtained in $\S$ 5.

Table 4 compares the band temperature estimates in the map basis and the
modulated data basis. The two sets of results agree to 0.5 $\sigma$. The
differences become larger for higher $\ell$ bins and possibly arise from 
non-circular beam effects. Souradeep \& Ratra (2001) show that non-circular 
beam effects become more important above the $\ell$ value corresponding to 
the inverse beamwidth.

\placetable{tbl4}

The Python V band powers can be used in combination with the results of
other experiments to test for consistency and constrain cosmological
parameters.  This can be done in a way that accounts for the
non--Gaussianity of the band power uncertainty by using the offset
log--normal form for the likelihood given in Bond, Jaffe, \& Knox (2000):  
\be
-2\ln{\cal L} = \sum_{a,b}\left(Z_a^{\rm t}-Z_a^{\rm d}\right)M_{ab}^Z\left(Z_b^{\rm t}-Z_b^{\rm d}\right)
\ee
where
\be
Z_a^{\rm d} \equiv  \ln(\C_a+x_a),
\ee
\be
Z_a^{\rm t} \equiv \ln\left(\sum_{l \in a}\C_l/N_l + x_a\right),
\ee
and
\be
M_{ab}^Z \equiv F_{ab}\left(\C_a+x_a\right)\left(\C_b+x_b\right).
\ee
Again $a, b$ denote bands. We have approximated the band--power
window function as a tophat with width $N_l = 30$.  

We have fit the $x_a$ and $\C_a$  parameters of this form to our
one--dimensional likelihood curves, as directly evaluated from the
modulated data in Table 4.  The Fisher matrix comes from the quadratic
estimator applied to the maps.  Table 5 gives the parameters of
the offset log-normal analytic fits to the band power likelihoods.

\placetable{tbl5}

Table 6 compares band temperatures estimated with and without the
circular beam approximation, from single modulation analyses where
the correlations between modulations are ignored
(as in Coble et al. 1999). The last column
corrects the results obtained by Coble et al. (1999) for a systematic
underestimation of the error bars by a factor of $\log_{10}e$. These results
also use the non-circular beam and do not use the flat-sky approximation 
(Souradeep \& Ratra 2001). The effect
of the circular beam approximation on the Python V power spectrum is minimal,
but would be greater for an experiment with higher S/N.

\placetable{tbl6}


\section{COMPARISON WITH PYTHON III}

Using the technique of $\S$ 4, maps of the sky are
constructed from the Python III data. We compared the maps in two different 
ways.
First, we decomposed each map into its signal to noise eigenmodes.
Keeping all the modes results in little useful
visual information since most of the features in such a map are noise.
Therefore, we excluded all modes with S/N
less than $1.7$; stopping at S/N $=1$ retains too
much noise. The resulting maps
are shown in Figure
\ref{fig7}. 
Since Python III has higher S/N than Python V, it retains many more modes.
Therefore, not all the features seen in the Python III
map should be visible in the
Python V map. However, structures found in
the Python V map are evident in the Python III map,
implying that Python III and Python V are consistent with each other.

\placefigure{fig7}

While the visual comparison is quite useful, it is difficult
to judge the significance of the agreement in this manner.
For a more quantitative comparison we use the
$\beta$ test of Knox et al. (1998).  This statistic
has a number of possible interpretations, one of
which is that it is the log of the ``probability enhancement
factor''.  That is, it tells us how much more probable
the data sets 1 and 2 are viewed jointly, as opposed to disjointly:
\be
\beta \equiv \ln{ P(T_1, T_2) \over P(T_1) P(T_2)}.
\ee
This can be re-written in terms of the likelihood function:
\be
\beta = \ln {\cal L}(T_1,T_2) - \ln {\cal L}(T_1) - \ln {\cal L}(T_2)
\ee
The joint likelihood for the two data sets uses the likelihood equation with the data
and the noise covariance being a concatenation of those from the two data sets.
It also uses the theory covariance between the two sets.

We find $\beta = 6.9$, which means that the data is $e^{6.9} = 992$ times
more probable viewed jointly than disjointly.  
This shows that the data sets have significantly more
in common than they would if they were unrelated to each other.

We can also examine how likely this value of $\beta$ is, under
these two different assumptions.  The first assumption
is that the data sets are related to each other exactly as we expect
due to their locations on the sky, and our inference of
the signal power spectrum from the Python V data.  With
assumption 1 we find that
$\langle \beta \rangle = 11.7 \pm 4.3$.  If instead
we assume that the two data sets are completely unrelated (perhaps
because one is completely contaminated), then 
$\langle \beta \rangle = -19.9 \pm 9.4$.  We
see that $\beta$ differs by 1.1 standard deviations from
the expected value and 2.8 standard
deviations from the value {\it expected in the absence of
cross-correlations}.

The $\beta$ statistic is model-dependent, but we found that
it changed by less than 15\% as we varied the amplitude
of the assumed power spectrum by amounts consistent with
the error bars and as we adjusted the calibration by $\pm 30\%$.


\section{CONCLUSIONS}

\begin{figure}
\epsscale{1.0}
\plotone{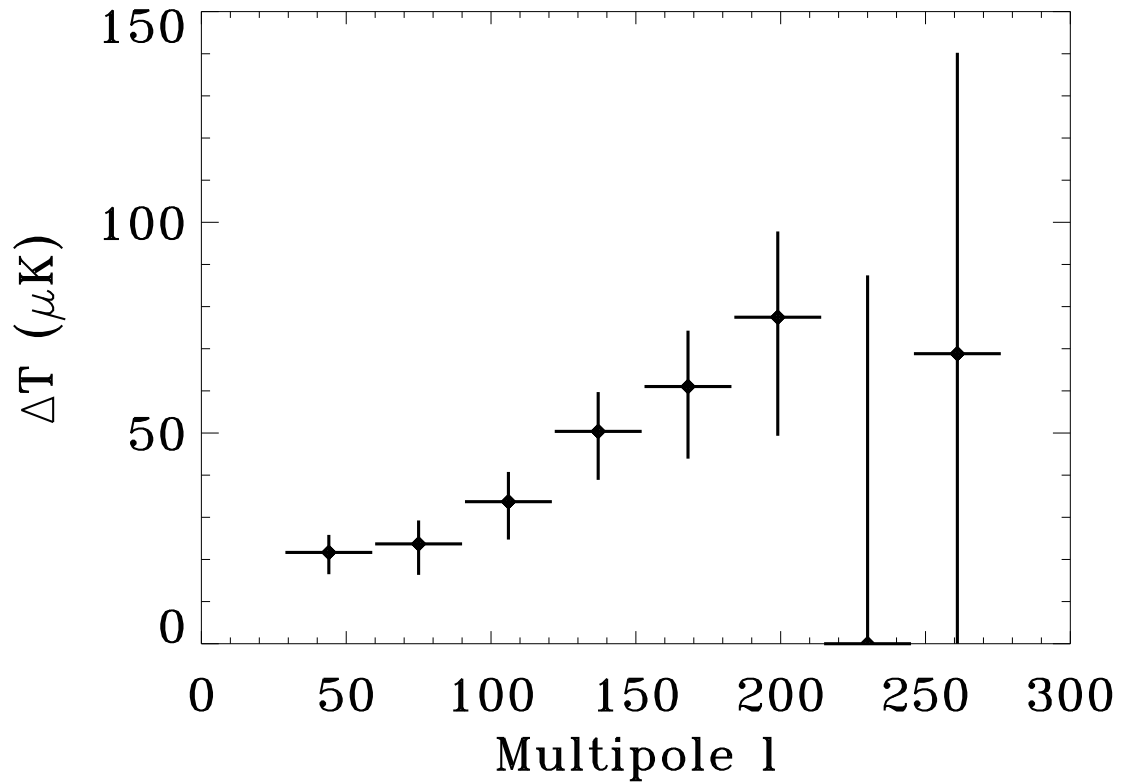}
\caption{Angular power spectrum from the
cross-modulation analysis. Band powers are
simultaneously estimated in 8 $\ell$-space bands.
Horizontal bars indicate the width of the bands.
Errors are $1\sigma$ statistical only and do not include calibration or beam
uncertainties. The last two bins do not show 2$\sigma$ detections.
The corresponding 2$\sigma$ upper limits on $\Delta T_\ell$ are
174 and 211 $\mu$K, respectively. 
\label{fig6}}
\end{figure}

\begin{figure}
\epsscale{1.0}
\plotone{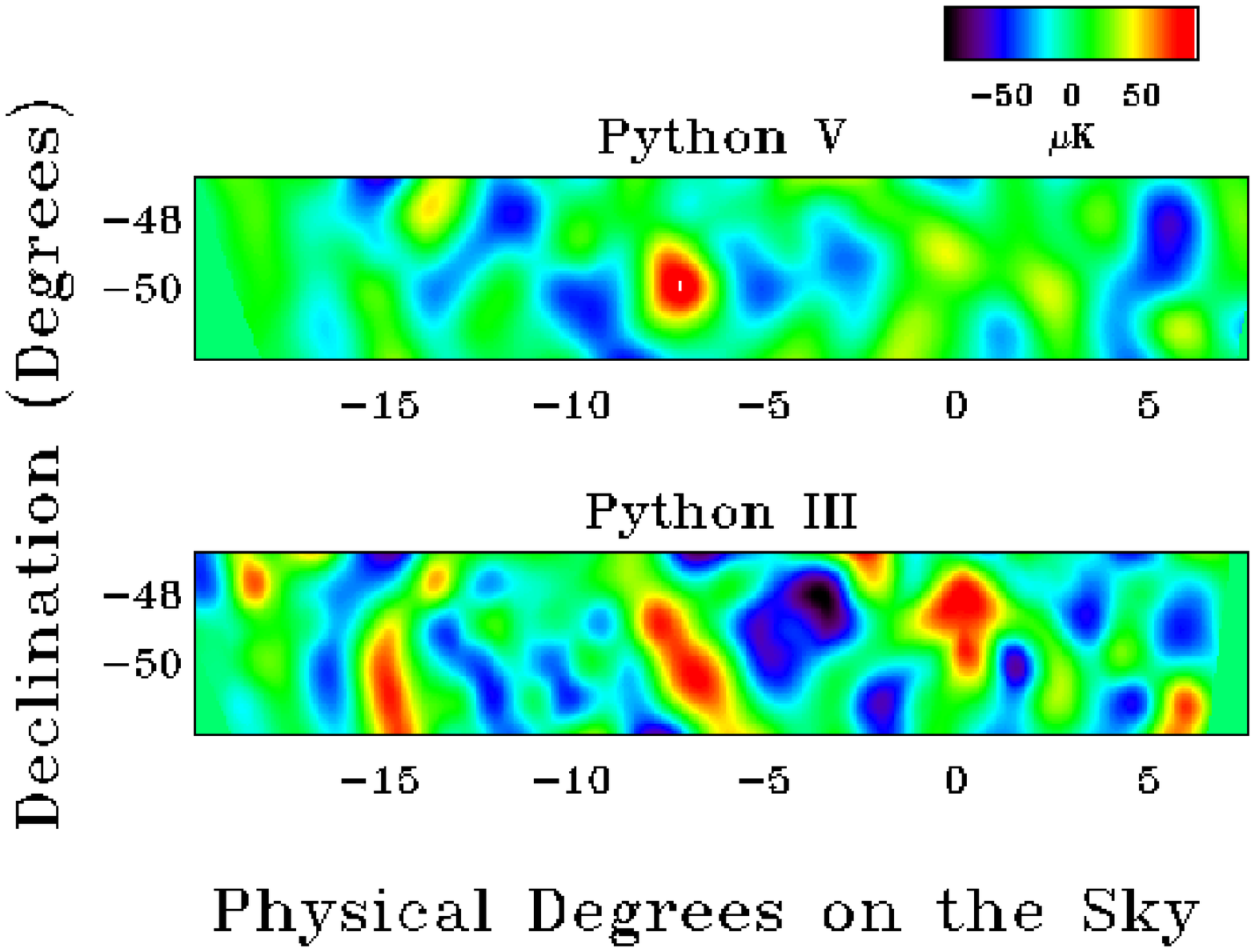}
\caption{Comparison of Python V and Python III maps.
The maps are made using only the highest S/N modes.
Structures found in the Python V map are evident in the Python III map,
implying that Python III and Python V are consistent with each other.
\label{fig7}}
\end{figure}

The Python V experiment densely samples 598 square degrees of the microwave 
sky and constrains the CMB anisotropy angular power spectrum from $\ell 
\sim$ 40 to $\ell \sim$ 260, showing that power is increasing from large
to smaller ($\ell \sim$ 200) angular scales. The noise matrix constructed in 
$\S$ 3 enables us to simultaneously estimate the angular power spectrum in
eight bands. The power spectra estimated from the map and directly from the 
modulated data are consistent. The rise seen in Figure \ref{fig6} 
is characteristic of acoustic oscillations in the early Universe.
A number of other measurements also indicate such a rise in power
(e.g., Ganga, Ratra, \& Sugiyama 1996; Netterfield et al. 1997;
de Oliveira-Costa et al. 1998; Torbet et al. 1999; Podariu et al. 2001; as well
as experiments mentioned in $\S$~1).
Python V extends to larger scales (lower $\ell$) than these,
to the smallest scales to which COBE was sensitive.

The Python III and V experiments differ in significant ways, including
frequency, receiver, year, and noise properties. Nevertheless, the maps 
and the $\beta$ test in $\S$ 7 indicate that they both detect similar 
signals, a rare and very valuable consistency check and confirmation.


\acknowledgments

This work was supported by the James S. McDonnell Foundation, PYI grant
NSF AST 90-57089, and the NSF under a cooperative agreement with the
Center for Astrophysical Research in Antarctica (CARA), grant NSF OPP
89-20223. CARA is an NSF Science and Technology Center. KC is supported
by NSF grant AST-0104465.
The work of SD was supported by the DOE and by NASA grant NAG 5-10842
at Fermilab, and by NSF grant PHY-0079251 at Chicago.
BR and TS acknowledge support from NSF CAREER grant AST-9875031.



\clearpage

\clearpage

\clearpage

\clearpage

\clearpage

\notetoeditor{Figure 5 should span both columns in the printed version}

\clearpage

\clearpage

\clearpage

\begin{table*}
\begin{center}
\begin{tabular}{|ll|}
\tableline
Term & Definition\\
\tableline
Field	&	Center position for an observation.\\
Cycle	&	One back-and-forth scan along the sky centered on one\\
	&	field. Consists of 128 samples.\\
Stare	&	164 consecutive cycles, again centered on the same field.\\
File	&	Approximately 10 consecutive stares. Each stare is\\
	&	centered on a new adjacent field.\\
Set	&	Approximately 100 consecutive files (about 13 hours of\\
	&	data taking) of the same $\sim$ 10 fields.\\
\tableline
\end{tabular}
\end{center}
\caption{Data subset terminology. See Figure \ref{fig1} for a visual representation of the sets of fields.}
\label{tbl1}
\end{table*}

\clearpage

\begin{table*}
\begin{center}
\begin{tabular}{|ccc|}
\tableline
Bin & $\ell$ & $\Delta T_\ell\ (\mu {\rm K})$\\
\tableline
1 &  $44^{+15}_{-15}$    &     $ 22_{-5}^{+4}$ \\
2 &  $75^{+15}_{-15}$    &     $ 24_{-7}^{+6}$ \\
3 &  $106^{+15}_{-15}$   &     $ 34_{-9}^{+7}$ \\
4 &  $137^{+15}_{-15}$   &     $ 50_{-12}^{+9}$ \\
5 &  $168^{+15}_{-15}$   &     $ 61_{-17}^{+13}$ \\
6 &  $199^{+15}_{-15}$   &     $ 77_{-28}^{+20}$ \\
7 &  $230^{+15}_{-15}$   &     $0^{+87}_{-0}$ \\
8 &  $261^{+15}_{-15}$   &     $ 69_{-69}^{+71}$  \\
\tableline
\end{tabular}
\end{center}
\caption{Python V final angular power spectrum. This comes from the 
cross-modulation analysis.
Note-- Band powers are simultaneously estimated in 8 $\ell$-space bands.
Error bars are 1$\sigma$ statistical only
and do not account for the calibration or beam uncertainties.
The last two bins do not show 2$\sigma$ detections. The corresponding 
2$\sigma$ upper limits on $\Delta T_\ell$ are 174 and 211 $\mu$K, 
respectively.}
\label{tbl2}
\end{table*}

\clearpage

\begin{table*}
\begin{center}
\begin{tabular}{|c|cccccccc|}
\tableline
 & 1 & 2 & 3 & 4 & 5 & 6 & 7 & 8\\
\tableline
1 & 1.00 & -0.164 & 0.012  & -0.008 &  -0.003 &  -0.003 &  -0.002 & -0.002 \\
2 & -0.164 & 1.00 & -0.211 & 0.019  & -0.011  &  -0.004 &  -0.002 & -0.004 \\
3 &  0.012 &  -0.211 & 1.00 & -0.217 & 0.024  &  -0.012 &  0.000 & -0.014 \\
4 & -0.008 & 0.019  & -0.217 & 1.00 & -0.228  &  0.030 & 0.005 & -0.064 \\
5 & -0.003 & -0.011 & 0.024 & -0.228 & 1.00   & -0.236 &    0.017 & -0.089 \\
6 & -0.003 & -0.004 & -0.012 & 0.030 & -0.236 & 1.00 & -0.361 & -0.003 \\
7 & -0.002 & -0.002 & 0.000  & 0.005 & 0.017  & -0.361 & 1.00 & -0.385 \\
8 & -0.002 & -0.004 & -0.014 & -0.064 & -0.089 & -0.003 & -0.385 & 1.00 \\
\tableline
\end{tabular}
\end{center}
\caption{Fisher Matrix $F^{-1}_{ab}/(\sqrt{F^{-1}_{aa}F^{-1}_{bb}})$.}
\label{tbl3}
\end{table*}

\clearpage

\begin{table*}
\begin{center}
\begin{tabular}{|cccc|}
\tableline 
$ {\rm Bin} $ & $ \ell $ & $ \Delta T_\ell\ (\mu {\rm K})$ 
  & $ \Delta T_\ell\ (\mu {\rm K})$ \\ 
        &          & (Map)                     & (Modulated Data) \\
\tableline
1 &  $44^{+15}_{-15}$    & $25^{+5}_{-6}$ 	  &  $25^{+6}_{-5}$ \\
2 &  $75^{+15}_{-15}$    & $23^{+5}_{-7}$ 	  &  $22^{+6}_{-5}$ \\
3 &  $106^{+15}_{-15}$   & $34^{+7}_{-8}$   	  &  $32^{+8}_{-7}$ \\
4 &  $137^{+15}_{-15}$   & $51^{+9}_{-11}$  	  & $50^{+9}_{-9}$ \\
5 &  $168^{+15}_{-15}$   & $60^{+12}_{-16}$ 	  & $56^{+15}_{-17}$\\
6 &  $199^{+15}_{-15}$   & $74^{+18}_{-25}$ 	  & $67^{+22}_{-20}$\\
7 &  $230^{+15}_{-15}$   & $0^{+56}_{-0}$          & $9^{+53}_{-9}$\\
8 &  $261^{+15}_{-15}$   & $46^{+78}_{-45}$   	  & $70^{+39}_{-70}$ \\
\tableline
\end{tabular}
\end{center}
\caption{Comparison of angular power spectra from the
map and from the modulated data.
Note-- These are from analyses that account for the cross-correlations
between modulations in a more limited manner than the full simultaneous
band power estimation of Table \ref{tbl2}, as described in the text.
Error bars are 1$\sigma$ statistical only.
The last two bins do not show 2$\sigma$ detections. The corresponding 
2$\sigma$ upper limits on $\Delta T_\ell$ are 112 and
202 $\mu {\rm K}$ (from the  map) and 104 and
181 $\mu {\rm K}$ (from the modulated data), respectively.}
\label{tbl4}
\end{table*}

\clearpage

\begin{table*}
\begin{center}
\begin{tabular}{|ccc|}
\tableline 
${\rm Bin}$	  &$x(\mu {\rm K}^2)$		& $\C_a (\mu {\rm K}^2)$\\
\tableline 
1		  & 100  	&620  \\
2     		  & 200	 	& 500  \\
3     		  & 500	 	& 1050 \\
4     		  &5000	 	& 2600 \\
5     		  &5000		& 3150\\
6     		  &6000 	&4570\\
7     		  &20000	& 0 \\
8      	 	  &60000	&5000\\
\tableline
\end{tabular}
\end{center}
\caption{Parameters of the offset log-normal analytic form
for the band power likelihood.}
\label{tbl5}
\end{table*}

\clearpage

\begin{table*}
\begin{center}
\begin{tabular}{|cccc|}
\tableline 
Modulation & $\ell$ & $\Delta T_\ell\ (\mu {\rm K})$
    & $\Delta T_\ell\ (\mu {\rm K})$ \\
  &  & (circular beam) & (elliptical beam) \\
\tableline
1 & $50^{+44}_{-29}$ &  $23^{+5}_{-4}$ 	 	 &  $23^{+5}_{-4}$ \\
2 & $74^{+56}_{-39}$ &  $24^{+6}_{-6}$ 	  	&   $24^{+6}_{-6}$ \\
3& $108^{+49}_{-41}$ &  $30^{+7}_{-7}$   	  & $30^{+7}_{-7}$ \\
4& $140^{+45}_{-41}$ &  $31^{+12}_{-13}$    	&   $30^{+12}_{-13}$ \\
5& $172^{+43}_{-40}$ &  $60^{+16}_{-17}$ 	  & $57^{+15}_{-16}$\\
6& $203^{+40}_{-38}$ &  $102^{+24}_{-24}$ 	  & $95^{+22}_{-22}$\\
7& $233^{+40}_{-38}$ &  $69^{+34}_{-64}$     	&   $61^{+30}_{-57}$\\
8& $264^{+39}_{-37}$ &  $0^{+90}_{-0}$	  	&   $0^{+78}_{-0}$ \\
\tableline
\end{tabular}
\end{center}
\caption{Comparison of single modulation angular power spectra with and without
the circular beam approximation.
Note-- These are from analyses that ignore 
correlations between the modulations. The last column
corrects the results obtained by Coble et al. (1999) for a systematic
underestimation of the error bars by a factor of $\log_{10}e$.
Error bars are 1$\sigma$ statistical only.
The last two bins do not show 2$\sigma$ detections. The corresponding 
2$\sigma$ upper limits on $\Delta T_\ell$ are 158 and
165 $\mu {\rm K}$ (for the  circular beam) and 142 and
143 $\mu {\rm K}$ (for the elliptical beam), respectively.}
\label{tbl6}
\end{table*}



\begin{thebibliography}{}


\bibitem[(Alvarez 1996)]{alv} Alvarez, D. 1996, PhD thesis, Princeton University

\bibitem[(Bond et al. 1998)]{bjkI} Bond, J. R., Jaffe, A., and Knox, L. 1998, Phys. Rev. D57, 2117

\bibitem[(Bond et al. 2000)]{bjkII} Bond, J. R., Jaffe, A., and Knox, L. 2000, ApJ, 533, 19

\bibitem[(Coble 1999)]{kc_thesis} Coble, K. 1999, PhD thesis, University of Chicago, {\tt astro-ph/9911419}

\bibitem[(Coble et al. 1999)]{py5_1} Coble, K., et al. 1999, ApJ, 519, L5

\bibitem[de Oliveira-Costa et al. (1998)]{deoliveiracosta98} de Oliveira-Costa, A., Devlin, M. J., Herbig, T., Miller, A. D., Netterfield, C. B., Page, L. A., \& Tegmark, M. 1998, \apj, 509, L77

\bibitem[Dragovan et al. 1994]{dragovan94} Dragovan, M., Ruhl, J. E., Novak, G., Platt, S. R., Crone, B., Pernic, R., \&\ Peterson, J. B. 1994 ApJ, 427, L67

\bibitem[Ganga et al. (2002)]{ganga01} Ganga, K., et al. 2002, in preparation

\bibitem[Ganga, Ratra, \& Sugiyama (1996)]{ganga96} Ganga, K., Ratra, B., \& Sugiyama, N. 1996, \apj, 461, L61

\bibitem[(Gundersen et al. 1995)]{jg} Gundersen, J. O., et al. 1995, \apjl, 443, L57

\bibitem[Halverson et al. (2002)]{dasi} Halverson, N. W., et al. 2002, \apj, 568, 38

\bibitem[(Knox et al. 1998)]{beta} Knox, L., Bond, J. R., Jaffe, A. H., Segal, M., \& Charbonneau, D. 1998 Phys. Rev. D, 58, 083004

\bibitem[(PyIV)]{py4} Kovac, J., Dragovan, M., Schleuning, D. A., Alvarez, D., Peterson, J. B., Miller, K., Platt, S. R., Novak, G. 1997, BAAS, 29.5, 112.04

\bibitem[(Maxima)]{maxima} Lee, A. T., et al. 2001, \apj, 561, L1


\bibitem[Miller et al. (1999)]{toco} Miller, A. D., et al. 1999, \apj, 524, L1 

\bibitem[(BOOMERANG)]{boom} Netterfield, C. B., et al. 2002, \apj, 571, 604

\bibitem[Netterfield et al. (1997)]{netterfield97}  Netterfield, C. B., Devlin, M. J., Jarosik, N., Page, L., \& Wollack, E. J. 1997, ApJ, 474, 47 

\bibitem[Park et al. (2001)]{park01} Park, C.-G., Park, C., Ratra, B., \& Tegmark, M. 2001, ApJ, 556, 582

\bibitem[Platt et al. 1997]{platt97} Platt, S. R., Kovac, J., Dragovan, M., 
  Peterson, J. B., \& Ruhl, J. E. 1997, ApJ, 475, L1 

\bibitem[Podariu et al. (2001)]{podariu01} Podariu, S., Souradeep, T., Gott, J. R., Ratra, B., \& Vogeley, M. S. 2001, ApJ, 559, 9


\bibitem[Rocha et al. (1999)]{rocha99} Rocha, G., Stompor, R., Ganga, K., Ratra, B., Platt, S. R., Sugiyama, N., \& G\'orski, K. M. 1999, ApJ, 525, 1

\bibitem[Ruhl et al. 1995]{ruhl95} Ruhl, J. E., Dragovan, M., Platt, S. R., Kovac, J., \&\ Novak, G. 1995, ApJ, 453, L1

\bibitem[Shandarin et al. (2002)]{shandarin01} Shandarin, S. F., Feldman,  H. A., Xu, Y., \& Tegmark, M. 2002, \apjs, 141, 1

\bibitem[Souradeep \& Ratra (2001)]{souradeep01} Souradeep, T., \& Ratra, B. 2001, \apj, 560, 28

\bibitem[(Tegmark 1997)]{teg97} Tegmark, M. 1997, Phys.Rev. D, 55, 5895

\bibitem[(Toco 1997)]{toco97} Torbet, E., et al. 1999, \apj, 521, L79

\bibitem[(Wilson et al. 2000)]{msamredo} Wilson, G. W., et al. 2000, ApJ, 532, 57

\bibitem[Wu et al. (2001)]{wu} Wu, J.-H. P., et al. 2001, Phys. Rev. Lett., 87, 251303

\end{thebibliography}
\end{document}